\documentclass[twocolumn,showpacs,prl]{revtex4}
\usepackage{graphicx}% Include figure files

\begin{document}

\title{Emergence of central mode in the paraelectric phase of ferroelectric perovskites }

\author{J. Weerasinghe$^{1}$, L. Bellaiche$^{1}$, T. Ostapchuk$^{2}$, P.
Kuzel$^{2}$, C. Kadlec$^{2}$, S. Lisenkov$^3$, I. Ponomareva$^3$,
and J. Hlinka$^{2}$} \email{hlinka@fzu.cz} \affiliation{ $^{1}$
Department of Physics and Institute for Nanoscience and
Engineering,
University of Arkansas, Fayetteville, Arkansas 72701, USA \\
$^{2}$ Institute of Physics, Academy of Sciences of the Czech
Republic, Na Slovance 2, 18221 Prague 8, Czech Republic\\
 $^{3}$
Department of Physics, University of South Florida, Tampa, Florida
33620, USA}

\begin{abstract}
{THz-range dielectric spectroscopy and first-principle-based
 effective-Hamiltonian molecular dynamics
simulations were employed to elucidate
 the dielectric response in the paraelectric phase of
(Ba,Sr)TiO$_3$ solid solutions. Analysis of the resulting
dielectric spectra suggests the existence of a crossover between
two different regimes: a higher-temperature regime governed by the
soft mode only {\it versus} a lower-temperature regime exhibiting
a coupled soft mode/central mode dynamics.
 Interestingly,
a single phenomenological coupling model can be used to adjust the
THz dielectric response in the entire range of the paraelectric
phase (i.e., even at high temperature). We conclude that
 the central peak is associated with thermally activated processes, and that it
 cannot be discerned anymore
 in the dielectric spectra when the rate of these
thermally activated processes exceeds certain characteristic
frequency of the system.}
 \end{abstract}

%77.22.Ch
%77.22.Gm: Dielectric loss and relaxation
%78.30.-j: Infrared and Raman spectra
%77.84.-s: Dielectric, piezoelectric, ferroelectric, and antiferroelectric materials.
%63.20.-e Phonons in crystal lattices.

 \maketitle
%1
It is well known that the static permittivity of ferroelectric
materials is related to frequencies of all polar phonon modes
through the Lyddane-Sachs-Teller formula \cite{LST}. Near the
phase transition, however, an additional low-frequency mode has
 to be often taken into account --
 the so-called central mode
\cite{JP,Axe,GY,Ono}. A generic reason for this additional
Debye-type excitation seems to be large-amplitude fluctuations
between quasi-stable off-center ionic positions. Existence of such
intrinsic central mode could be very clearly demonstrated, e.g.,
below the cubic-tetragonal phase transition $T_{\rm C}$ of
BaTiO$_3$ \cite{HlinkaPRL}.

%2
Similar central mode (CM) is also known to exist in the {\it
paraelectric phase}. Phenomenological theories of the paraelectric
CM have been developed by several authors \cite{Axe,JP,GY,PB,Ono}.
All these approaches lead to a coupled relaxator-oscillator
dielectric response. However,
  an important question has been left open so far: whether the CM
 persists up to the highest temperatures, or rather it progressively disappears, or whether it disappears at some
 well-defined temperature $T_{\rm CM} (> T_{\rm C})$.

%3
Unfortunately, it is much more difficult
 to obtain a clear-cut
experimental evidence for the dielectric CM in the cubic
perovskite phase \cite{JP,Vogt,Inna,GY}.
 The characteristic
frequencies of the soft phonon-oscillator and CM in KNbO$_3$ and
BaTiO$_3$ are so broad and close together that they can hardly be
disentangled. The difficulty of the experimental analysis of the
dielectric spectra of KNbO$_3$ and BaTiO$_3$ is caused, at least
partly, by intrinsically large phonon damping factors, related to
the fact that $T_{\rm C}$ is quite high in these compounds (about
700 and 400\,K, respectively). Therefore, it is of interest to
study the CM also in the ferroelectric compounds in which the
phase transition occurs at lower temperatures, which can be
achieved for example by mixing BaTiO$_3$ and KNbO$_3$ with
suitable incipient ferroelectrics, namely SrTiO$_3$ and KTaO$_3$,
respectively.

%4
Here, we describe a combined experimental and theoretical study
 of the technologically-relevant mixed Ba$_x$Sr$_{1-x}$TiO$_3$ system (BST) aimed to
establish the characteristic temperature trends in the model
parameters of the relaxator-oscillator dielectric response in the
polar perovskites. It turns out that this phenomenological theory
allows to ``easily'' understand the existence of the temperature
$T_{\rm CM}$, at which the CM in the dielectric spectra is
appearing or disappearing.

\begin{figure}
\includegraphics[width=7cm]{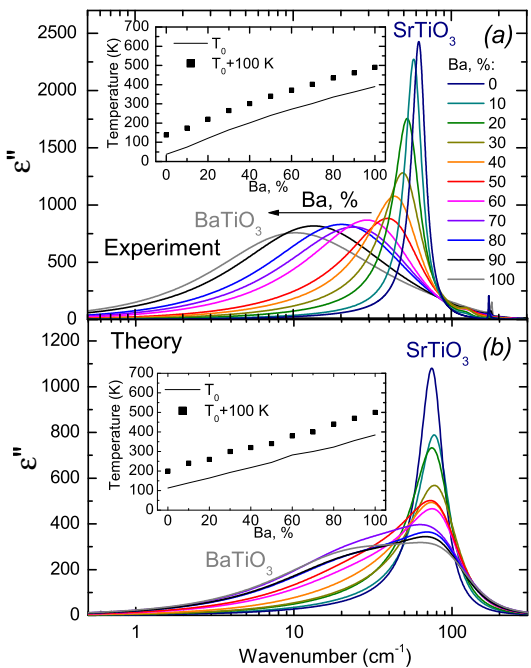}
\caption{(Color online) Dielectric loss spectra of BST ceramics
determined (a) by combined IR reflectivity and time-domain THz
transmission spectroscopic techniques and (b) by MD simulations
using the effective Hamiltonian of Ref. [\onlinecite{effH}]. The
spectra are obtained at 100\,K above the extrapolation temperature
$T_0$ of the Curie-Weiss law, indicated in the insets by full
line. Point symbols in insets indicates temperature-concentration
coordinates associated with the spectra shown in the main panel.}
\end{figure}

%5
Basic dielectric properties of BST solid solutions and the
concentration diagram of this system were thoroughly discussed,
see e.g. Refs.\,\onlinecite{Shirokov,effH} and references therein.
Experiments discussed below were carried out with a set of
high-density BST ceramics with Ba:Sr ratio ranging from x=0 to
x=1, prepared by methods described elsewhere (see
Refs.\,\cite{Tana1,Tana2}). Low-frequency permittivity obtained
from standard dielectric measurements (10 kHz) in the paraelectric
phase was fitted to a Curie-Weiss law
\begin{equation}
\varepsilon(0)^{-1} = (T-T_0)/C ~.\label{eq:C}
\end{equation}
As expected \cite{JP,Shirokov}, Curie constants were of the order
of $10^5$\,K for all concentrations, while the extrapolation
temperature $T_0(x)$ turned out to be strongly dependent on the
concentration (see the inset of Fig.\,1a).

%6
These findings may suggest that at a fixed temperature interval
above the extrapolation temperature $T_0$, the {\it static}
permittivity is roughly independent of $x$. However, this is not
the case for the THz-range dielectric spectra of BST ceramics. In
Fig.\,1a, we compare dielectric loss spectra
 obtained similarly as in
Refs.\,\onlinecite{Tana1,Tana2} from simultaneous fits to
time-domain THz transmission and far-IR reflectivity spectra.
Although all these spectra are obtained at 100\,K above the
corresponding temperature $T_0$, we can see a clear crossover
between two rather distinct regimes: a broad-band regime for
BaTiO$_3$-rich compounds, and a narrow-band regime for
SrTiO$_3$-rich compounds. The narrow response band can be easily
adjusted with a {\it single} damped harmonic oscillator (DHO)
model, while the broad band suggests a more complex spectral
shape, such as that of the coupled relaxator-oscillator model.

%7
In order to prove the intrinsic origin of this spectral broadening
phenomenon, we have also calculated \cite{effHnon,VCA1,VCA2}
dielectric spectra of disordered BST solutions by conducting
molecular dynamics (MD) simulations using the effective
Hamiltonians of Ref. \onlinecite{effH} -- as previously reported
in Refs.\,\onlinecite{Inna,HlinkaPRL}.
%NEW TEXT HERE
Chemical disorder in the Ba and Sr ion sublattice has been
included directly by specifying randomly chosen Ba and Sr ion
positions within a simulation supercell representing a 12x12x12
perovskite formula units of BST, assuming periodic boundary
conditions.
%Further details of these simulations are provided in the supplementary material \cite{epaps}.
 To verify that the
results are independent of the particular configuration,
simulations were also carried out for several different
configurations.

The resulting spectra (see Fig.1\,b) are quite monotonously
varying with the average Ba concentration in the simulated
supercell, even though the Ba ion distribution is obviously
different for each concentration. This also confirms that the
observed trends are driven by the overall concentration, rather
than by the degree of the occupational disorder.
 As consistent with the experiments,
the predicted static dielectric permittivity above $T_{\rm C}$
obeyed the Curie-Weiss law (Eq.\,1). Moreover, simulated loss
spectra for temperatures about 100\,K above $T_0$ indeed confirmed
a similar broadening for BaTiO$_3$-rich compounds as in the
measurements (see Fig.1\,b).

\begin{figure}
\includegraphics[width=6cm]{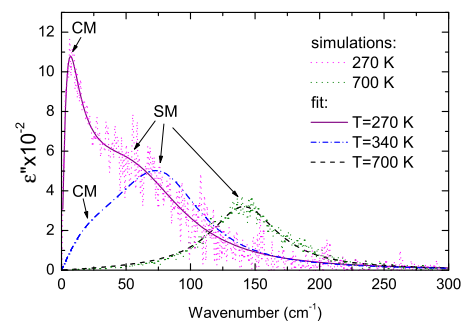}
\caption{(Color online) Dielectric loss spectra of
Ba$_{0.5}$Sr$_{0.5}$TiO$_3$ determined by MD simulations
corresponding to $T=270$\,K (at 25\,K above the theoretical
extrapolation temperature $T_0$), at $T=345$\,K and at $T=700$\,K.
Smooth continuous lines are fit to the model of coupled
oscillator-relaxator model of Eq.\,\ref{eq:3} (fitting parameters
are shown in Fig.\,3.)}
\end{figure}

Despite of this considerable broadening, spectra of Fig.\,1
do not show a very clear 2-maximum dielectric loss response,
such as the one observed in the A$_1$ spectra of the tetragonal
phase of BaTiO$_3$ single crystal \cite{HlinkaPRL}.
 We expect that CM and soft mode (SM) components are much better separated
  in the spectra within the first few tens of K above $T_{\rm C}$. However, in this temperature range the
THz measurements were not feasible even with very thin ceramic
samples because they become too opaque for THz radiation.

Motivated by this difficulty, we have calculated dielectric
spectra of Ba$_{0.5}$Sr$_{0.5}$TiO$_{3}$ (BST50) in a broad range
of frequencies and temperatures down to the vicinity of the
ferroelectric phase transition using the effective-Hamiltonian
approach of Ref. \onlinecite{effH}. In order to obtain
well-behaved smooth dielectric spectra especially closer to
$T_{0}$ in the low THz regime, MD time-step was chosen to be
0.5\,fs and the simulations were conducted for a length of
13.2\,ns in order to obtain 10000 samples for averaging the
autocorrelation functions \cite{Inna}.
 Figure \,2 displays
the frequency dependence of the imaginary part of dielectric
permittivity derived from MD simulations at 270\,K ($C = 0.6
\times 10^5$\,K and $T_{0}= 245$\,K for simulated BST50). The
presence of a lower frequency Debye-like CM in addition to the
higher frequency SM is clearly seen there.

To account for the simultaneously appearing SM and CM, we have used
the coupled oscillator-relaxator model (following the notation of
Ref.\,\onlinecite{HlinkaPRL}):
\begin{equation}
\varepsilon\left(\omega\right)= \varepsilon_{\infty}+
\frac{\Omega_{\rm S}^{2}}{\omega_{\rm S}^{2}-\omega^{2}
-i\omega\Gamma_{\rm S}-\frac{\gamma_{\rm
D}\delta^{2}}{(\gamma_{\rm D}-i\omega)}} ~.\label{eq:3}
\end{equation}
Here $\Omega_{\rm S}$, $\omega_{\rm S}$, $\Gamma_{\rm S}$ are the
plasma frequency, oscillator frequency and damping constant of the
SM, respectively; $\gamma_{\rm D}$ is the bare relaxation
frequency of the CM, $\delta$ is the coupling coefficient, and
$\varepsilon_{\infty}=1$ is the background permittivity
\cite{note13}. Model parameters of the unconstrained fitting of
Eq.\,\ref{eq:3} to the MD simulation data at various temperatures
are shown as full symbols in Fig.\,3. For temperatures above $\sim
400$\,K, the CM is not visible in the simulated spectrum and the
simple DHO formula (Eq.\,\ref{eq:3} with $\delta=0$) provides a
satisfactory fit (see Fig.\,2).

\begin{figure}
\includegraphics[width=7cm]{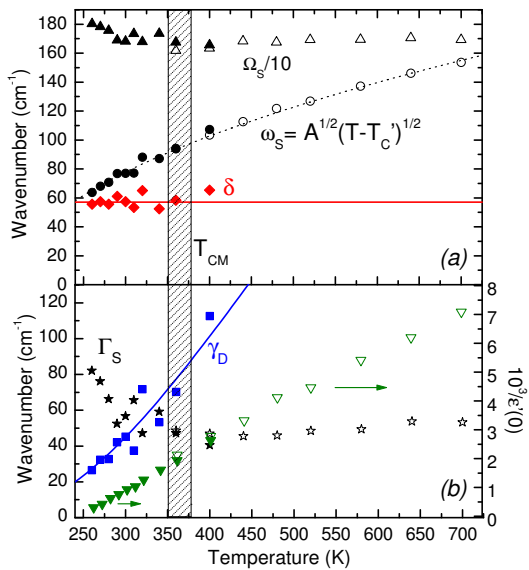}
\caption{(Color online) Temperature dependence of the parameters
 of Eq.\,2 obtained from MD dielectric spectra of
BST50.
   Filled symbols correspond to the unconstrained
  fit by Eq.\,\ref{eq:3} for temperatures below 400\,K,
full line in panel (b) stands for adjustment of the Arrhenius law
  (Eq.\,\ref{eq:4}) to  $\gamma_{\rm D}$ data below 400\,K.
  Open symbols were obtained  for temperatures above T$_{\rm CM}$
  from a {\it constrained} fit  by Eq.\,\ref{eq:3} with $\delta =
  57$\,cm$^{-1}$ and $\gamma_{\rm D}$ prescribed by Eq.\,\ref{eq:4} with
  $U=84$\,meV and $\gamma_{\infty}=1160$\,cm$^{-1}$.}
\end{figure}

What can we learn from the temperature trends revealed in Fig.\,3?
First, let us note that Eq.\,\ref{eq:3} can be considered as an
asymptotic case of the dielectric response in the model of two
coupled DHO's, one representing the normal phonon SM, and the
other representing the extraneous excitation D \cite{JP}:
\begin{eqnarray}
\varepsilon\left(\omega\right)= \varepsilon_{\infty}+ \left[
\begin{array}{cc} \Omega_{\rm D} &
\Omega_{\rm S}\end{array}
 \right]
{\bf \rm \hat{G}}(\omega)
 \left[\begin{array}{c}
\Omega_{\rm D}\\
\Omega_{\rm S}
\end{array}\right]~,
\label{eq:oGo}
\end{eqnarray}
with
\begin{eqnarray}
{\bf \rm \hat{G}}(\omega)^{-1} = \left[
\begin{array}{cc}
\omega_{\rm D}^{2}-\omega^{2}-i\omega\Gamma_{\rm D} & \Delta\\
\Delta^* & \omega_{\rm S}^{2}-\omega^{2}-i\omega\Gamma_{\rm S}
\end{array}\right] ~,
\label{eq:GDHO}
\end{eqnarray}
where the quantities $\Omega_{\rm D}$, $\omega_{\rm D}$,
$\Gamma_{\rm D}$ related to the CM have analogous meaning as
$\Omega_{\rm S}$, $\omega_{\rm S}$, $\Gamma_{\rm S}$. The coupling
term $\Delta$ can be chosen positive.
 Indeed, in the Debye-relaxator limit ($\omega \ll
\Gamma_{\rm D}$ and $\omega \ll \omega_{\rm D}$) the $\omega^2$
term can be neglected against $ \omega_{\rm
D}^{2}-i\omega\Gamma_{\rm D}$ and the above matrix reduces to that
of the coupled relaxator-oscillator model \cite{JP}.
 In this case, four parameters $\Omega_{\rm D}$, $\omega_{\rm D}$,
$\Gamma_{\rm D}$ and $\Delta$ can be replaced by only three
parameters - the relaxational frequency $\gamma_{\rm D}$, the
coupling frequency $\delta$, and an auxiliary frequency $\rho$
\begin{eqnarray}
\gamma_{\rm D} = \omega_{\rm D}^2/\Gamma_{\rm D}, ~ \delta =
\Delta/\omega_{\rm D},~ \rho=\omega_{\rm D} \Omega_{\rm
S}/\Omega_{\rm D} ~. \label{eq:Substi}
\end{eqnarray}
With these substitutions (Eq.\,\ref{eq:Substi}) and for
 $\omega \ll \Gamma_{\rm D}$, $\omega \ll
\omega_{\rm D}$, and $\Omega_{\rm S} \gg \Omega_{\rm D}$ or $\rho
\gg \delta$, the expression
 of Eq.\,\ref{eq:oGo} reduces to the formula of Eq.\,\ref{eq:3}.

%{\bf Anticipated temperature trends.}
  Previously, several authors \cite{JP,GY,PB}
  treated the problem of paraelectric CM
in perovskites and derived a frequency dependence of the
dielectric permittivity in the form equivalent to eqs.\,\ref{eq:3}
or \ref{eq:oGo}--\ref{eq:GDHO}.
 Common assumption in all these approaches \cite{JP,GY,PB} is
that the Curie-Weiss behavior of the static permittivity results
from the Cochran-like temperature dependence of the SM frequency
$\omega_{\rm S}^2 = A (T-T_{\rm C}')$ with a "Cochran" temperature
$T_{\rm C}' < T_{0}\sim T_{\rm C}$. This is indeed well obeyed
here (with $A \doteq 50$\,cm$^{-2}$K$^{-1}$ and $T_{\rm C}' =
165$\,K for BST50, see Fig.\,3). However, the implications for the
temperature dependence of $\delta$ and $\gamma_{\rm D}$ are quite
different among these models. In the spirit of
Ref.\,\onlinecite{JP}, one could assume that $\delta$ and
$\gamma_{\rm D}$ either do not change with temperature since the
bare D-mode is not contributing to the ordering mechanism, or
perhaps, that (model A)
\begin{equation}
\delta^2= {\rm const}.
\end{equation}
and
\begin{equation}
 \gamma_{\rm D} \sim 1/T,
\end{equation}
when assuming in Eq.\,(\ref{eq:Substi}) that the D-mode damping
increases with the temperature like $\Gamma_{\rm D} \sim T$. Bare
pseudospins in the model of Ref.\,\onlinecite{GY} are considered
as freely fluctuating bistable dipoles forming an ideal
paraelectric gas with Arrhenius relaxation law, which leads to
(model B)
\begin{equation}
\delta^2=\alpha_0/T
\end{equation}
and
\begin{equation}
\gamma_{\rm D}=\gamma_{\infty}\exp\left[-\frac{U}{kT}\right] ~,
\label{eq:4}
\end{equation}
where $U$ is a potential barrier and $\gamma_{\infty}$ is an
attempt frequency. In a similar model applied to BaTiO$_3$ in
Ref.\,\onlinecite{PB}, one assumes that bare pseudospin subsystem
would order at a finite temperature $T_{\rm L}< T_{\rm C}$, with
$T_{\rm L}\approx T_{\rm C}$ and the resulting formulas for
$\delta$ and $\gamma_{\rm D}$ read (model C)
\begin{equation}
\delta^2=\alpha_0/(T-T_{\rm L})
\end{equation}
and
\begin{equation}
\gamma_{\rm
D}=\gamma_{\infty}\exp\left[-\frac{U}{kT}\right]\frac{T-T_{\rm
L}}{T} ~.
\end{equation}

%{\bf Temperature trends in simulations.}
Comparison of the anticipated temperature trends with the outcome
of our MD simulations shown in Fig.\,3 indicates that none of the
above three models is fully satisfactory. As a matter of fact, the
MD results suggest that the coupling term $\delta$ is constant (or
it perhaps slightly increases with the increasing temperature),
which is compatible with the model A only, while the increasing
$\gamma_{\rm D}$ rather supports the thermally activated flipping,
assumed in models B or C. Tentatively, we have fitted the
temperature dependence of $\gamma_{\rm D}$ to the Arrhenius law of
Eq.\,\ref{eq:4} in the 260-400\,K temperature range. It gives
reasonable values of $U=84$\,meV and
$\gamma_{\infty}=1160$\,cm$^{-1}$.

In fact, it turns out that the BST50 dielectric spectra can be
 well fitted at {\it all} temperatures (i.e., even above $T_{\rm CM}$!)
 when keeping $\delta = 57$\,cm$^{-1}$ and  $\gamma_{\rm
D}$ described by the aforementioned Arrhenius law (fitted
parameters are shown by open symbols in Fig.\,3). In other words,
the coupled relaxator-oscillator model (Eq.\,\ref{eq:3}) can be
used in the whole temperature range above $T_{\rm C}$, without any
particular discontinuity in the temperature course of the model
parameters near the temperature $T_{\rm CM}$.

So why does the CM seem to "disappear" from the BST50 spectrum at
$T_{\rm CM}$? In order to understand this behavior, it is
convenient to consider the formula of Eq.\,\ref{eq:3} as the
response of a DHO with a frequency-dependent self-energy term
\begin{eqnarray}
{ \hat{\Pi}}(\omega) = -\frac{\delta^2\gamma_{\rm D}^2}{\gamma_{\rm
D}^2+\omega^2} -i\omega\ \left[ \Gamma_{\rm S} +
\frac{\delta^2\gamma_{\rm D}}{\gamma_{\rm D}^2+\omega^2} \right] ~.
\label{eq:PI}
\end{eqnarray}

From this expression, it is obvious that the simple DHO formula is
recovered when
 $ \gamma_{\rm
D} \gg \delta^2/ \Gamma_{\rm S}$, since in this case the imaginary
part of the self-energy becomes simply $Im[{ \hat{\Pi}}] \doteq
-\omega \Gamma_{\rm S}$.
 Therefore, it is reasonable to
 define the temperature $T_{\rm CM}$ as the temperature at which $\delta^2 \approx \gamma_{\rm D} \Gamma_{\rm
 S}$. Since the temperature dependence of $\gamma_{\rm D}$ is
 steep, the temperature-driven crossover between $ \gamma_{\rm
D} \ll \delta^2/ \Gamma_{\rm S} $ and $ \gamma_{\rm D} \gg
\delta^2/ \Gamma_{\rm S} $ is quite sharp.
%NEW TEXT
With the data of Fig.\,3, this crossover temperature is about
$T_{\rm CM}\approx 360-370\,K$. Moreover, in case of Arrhenius law
for $\gamma_{\rm
 D}$,
  one obtains
  \begin{equation}
  T_{\rm CM}\approx \frac{U}{k \ln(\gamma_{\infty}\Gamma_{\rm
  S}/\delta^2)}~,
  \end{equation}
showing that $T_{\rm CM}$ scales with the activation energy $U$.

This kind of crossover temperature $T_{\rm CM}$ likely bears some
resemblance with the so-called Burns temperature and $T^*$
temperature, that have been observed in some complex perovskite
materials \cite{DulkinBTO,DulkinPMN,Burns}, since they all may be
associated with change of dynamics.

%\section{CONCLUSION}

In summary, we provide interpretation of experiments and MD
simulations suggesting that permittivity of BST system in the
paraelectric phase shows a crossover between a high-temperature
spectrum with a simple SM and a lower-temperature spectrum with a
more complex shape involving an additional CM that is coupled to
the SM. Analysis of MD simulations allowed us to select convenient
formula for the temperature evolution of the model parameters,
leading to a deviation with respect to previous theoretical
predictions when CM and SM are coupled. As a matter of fact, these
MD results indicate a thermally activated dynamics in a postulated
pseudospin subsystem and their almost temperature-independent
pseudospin-phonon coupling coefficient. Furthermore and
interestingly,
 the dielectric spectra of BST50 from MD simulations could also
 be analyzed using this original coupled CM-SM model at {\it any}
  temperature above $T_{\rm C}$ (i.e., even in the high-temperature regime).
  As a result, the appearance of the CM in the dielectric spectra
   is understood as a crossover between fast and slow pseudospin dynamics.
   Arrhenius-type temperature
dependence of bare pseudospin dynamics allows to understand the
abrupt change of the spectrum at this crossover, which thus
appears almost as a phase transition.

\begin{acknowledgments}
We thank Jan Petzelt for useful discussions. J.H.
%, T.O., P.K., and L.B.
acknowledge the support of the Czech Ministry of Education
(Project MSMT ME08109). J.W. and L.B. acknowledge the financial
support of NSF DMR-1066158 and DMR-0701558. They also acknowledge
ONR Grants N00014-11-1-0384 and N00014-08-1-0915, the Department
of Energy, Office of Basic Energy Sciences, under contract
ER-46612, and ARO Grant W911NF-12-1-0085 for discussions with
scientists sponsored by these grants. I.P. acknowledges the
financial support of the Department of Energy, Office of Basic
Energy Sciences under grant DE-SC0005245. Some computations were
also made possible thanks to the MRI grant 0722625 from NSF, ONR
grant N00014-07-1-0825 (DURIP), and a Challenge grant from the
Department of Defense.
\end{acknowledgments}


\begin{thebibliography}{99}
\bibitem{note13} Similar spectra, as e.g. those of Ref.\,\onlinecite{Tana2} are often fitted
 to a sum of two independent DHO's. It is important to note that the present model
 has less parameters and still gives a very good description of the data.
\end{thebibliography}
\end{document}